\documentclass[british,english]{amsart}
\usepackage[T1]{fontenc}
\usepackage[latin9]{inputenc}
\usepackage{babel}
\usepackage{verbatim}
\usepackage{enumitem}
\usepackage{amstext}
\usepackage{amsthm}
\usepackage{amssymb}
\usepackage[pdfusetitle,
 bookmarks=true,bookmarksnumbered=false,bookmarksopen=false,
 breaklinks=false,pdfborder={0 0 1},backref=false,colorlinks=false]
 {hyperref}

\makeatletter
\numberwithin{equation}{section}
\numberwithin{figure}{section}
\theoremstyle{plain}
\newtheorem{thm}{\protect\theoremname}[section]
\theoremstyle{definition}
\newtheorem{example}[thm]{\protect\examplename}
\newlist{casenv}{enumerate}{4}
\setlist[casenv]{leftmargin=*,align=left,widest={iiii}}
\setlist[casenv,1]{label={{\itshape\ \casename} \arabic*.},ref=\arabic*}
\setlist[casenv,2]{label={{\itshape\ \casename} \roman*.},ref=\roman*}
\setlist[casenv,3]{label={{\itshape\ \casename\ \alph*.}},ref=\alph*}
\setlist[casenv,4]{label={{\itshape\ \casename} \arabic*.},ref=\arabic*}

\usepackage{upgreek}
\usepackage{amssymb}
\let\originalleft\left
\let\originalright\right
\renewcommand{\left}{\mathopen{}\mathclose\bgroup\originalleft}
\renewcommand{\right}{\aftergroup\egroup\originalright}

\makeatother

\addto\captionsbritish{\renewcommand{\casename}{Case}}
\addto\captionsbritish{\renewcommand{\examplename}{Example}}
\addto\captionsbritish{\renewcommand{\theoremname}{Theorem}}
\addto\captionsenglish{\renewcommand{\casename}{Case}}
\addto\captionsenglish{\renewcommand{\examplename}{Example}}
\addto\captionsenglish{\renewcommand{\theoremname}{Theorem}}
\providecommand{\casename}{Case}
\providecommand{\examplename}{Example}
\providecommand{\theoremname}{Theorem}

\begin{document}
\title[Periods of $N$-Body Systems]{Periods of $N$-Body Systems\\
Determined through Dimensional Analysis }
\author{Dan Jonsson}
\begin{abstract}
A generalization of classical dimensional analysis, presented in a
separate article, makes it possible to derive Kepler's third law for
the period of a two-body system, up to a multiplicative constant,
without solving the equations of motion. Here we show how to derive
generalizations of Kepler's third law to $n$-body systems by the
same technique. Our results agree with conjectures by Sun on the period
of a classical $n$-body system and by Semay and Sun on the quantum-theoretical
counterpart of the period of a classical $n$-body system.
\end{abstract}

\maketitle

\section{\label{sec:1}Introduction}

The classical theory of $n$-body systems concerns $n$ point masses,
``bodies'' or ``particles'' $p_{1},\ldots,p_{n}$ that move in
space under influence of their mutual gravitational attraction in
accordance with the laws of Newtonian mechanics. The equations of
motion for a collisionless $n$-body system can be derived directly
from the force law $\mathbf{F}=m\mathbf{a}$ and Newton's law of gravitation
$\mathbf{F}=G\frac{m_{1}m_{2}}{r^{2}}\mathbf{r}$, where $\mathbf{r}$
is a unit vector pointing in the same direction as $\mathbf{F}$.
In vector form, the equations are
\begin{equation}
\begin{cases}
m_{i}\frac{\mathrm{d}^{2}\mathbf{q}_{i}}{\mathrm{d}t^{2}}=G\sum_{j\in\mathbf{n}\setminus\left\{ i\right\} }m_{i}m_{j}\frac{\left(\mathbf{q}_{j}-\mathbf{q}_{i}\right)}{\left\Vert \mathbf{q}_{j}-\mathbf{q}_{i}\right\Vert ^{3}} & \left(i=1,\ldots,n\right),\end{cases}\label{eq:motion}
\end{equation}
where $\mathbf{n}=\left\{ 1,\ldots,n\right\} $, $t$ is time, $G$
the gravitational constant, $m_{i}$ the mass of $p_{i}$ and $\mathbf{q}_{i}$
its radius vector as a function of $t$.

For a spatially bounded $n$-body system there may exist some time
$T_{1}$ such that at a later time $T_{1}+T$ each body has the same
position and velocity as at $T_{1}$. Then it has the same position
and velocity at $T_{k+1}=T_{1}+kT$ for any positive integer $k$
since the equations of motion determine the future trajectory of each
body given only its position and velocity at an arbitrary time. We
call a trajectory followed from $T_{k}$ to $T_{k+1}$ an \emph{orbit},
an $n$-body system with trajectories made up of orbits an \emph{orbital
(periodic)} $n$-body system, and $T$ the \emph{period} of the orbital
system.

In the two-body case, we obtain, under suitable initial conditions,
the familiar elliptical orbits, with the two bodies in addition obeying
Kepler's second and third laws. Kepler's third law concerns the period
of an orbital two-body system. It can be shown from the equations
of motion that
\begin{equation}
T^{2}=4\pi^{2}a^{3}G^{-1}\left(m_{1}+m_{2}\right)^{-1},\label{eq:period-2-1}
\end{equation}
where $T$ is the period and $a$ the semi-major axis of the elliptical
orbit \cite[eq. 2.37]{CAR}. 

In a two-body system, the semi-major axis $a$ of the ellipse describing
the orbit of $p_{1}$ about $p_{2}$, whose radius vector is set to
$\mathbf{0}$, is the mean extremal distance $\mu_{12}$ given by
\begin{equation}
\mu_{12}=\frac{\left\Vert \mathbf{r}_{a}-\mathbf{0}\right\Vert +\left\Vert \mathbf{r}_{p}-\mathbf{0}\right\Vert }{2},\label{eq:a-1}
\end{equation}
where $\mathbf{r}_{a}$ and $\mathbf{r}_{p}$ are the radius vectors
of the apoapsis and periapsis, respectively, of $p_{1}$ relative
to $p_{2}$.%
\begin{comment}
Every two-body system may be regarded as a system of two bodies each
of which travels around the center of mass ($\mathrm{CM}$) of the
system in an elliptical orbit, and denoting the radius vector of $\mathrm{CM}$
by $\mathbf{r}{}_{\mathrm{CM}}$ we also have
\begin{equation}
a=\frac{\left\Vert \mathbf{r}_{1a}-\mathbf{r}{}_{\mathrm{CM}}\right\Vert +\left\Vert \mathbf{r}_{1p}-\mathbf{r}{}_{\mathrm{CM}}\right\Vert +\left\Vert \mathbf{r}_{2a}-\mathbf{r}{}_{\mathrm{CM}}\right\Vert +\left\Vert \mathbf{r}_{2p}-\mathbf{r}{}_{\mathrm{CM}}\right\Vert }{2},\label{eq:a-2}
\end{equation}
where $\mathbf{r}_{ia}$ and $\mathbf{r}_{ip}$ are the radius vectors
of the apoapsis and periapsis, respectively, of $p_{i}$ relative
to $\mathrm{CM}$.
\end{comment}

The semi-major axis $a$ is a \emph{characteristic distance}, that
is, $T$ depends separately on the mean extremal distance between
the bodies (up to a multiplicative constant). It turns out that $T$
depends on any $d\not\propto a$ (for example, the semi-minor axis)
only in conjunction with a dimensionless shape parameter $s$ (such
as the eccentricity of the orbital ellipse), so that $T^{\mathfrak{K}}=\varphi\left(m_{1},m_{2},a,G\right)$
becomes $T^{\mathfrak{K'}}=\varphi'\left(m_{1},m_{2},d,s,G\right)$.

Let $E$ be the total mechanical energy of an orbital $n$-body system
in a center-of-mass inertial frame, and let the potential energy tend
to $0$ from below as the distances between bodies approach infinity.
(These conditions ensure that the sign of $E$ is meaningful.) Then
it can be shown that in the two-body case we have
\begin{equation}
E=-\frac{Gm_{1}m_{2}}{2a}.\label{eq:121}
\end{equation}
Here, $E<0$ since otherwise $a<0$. (\ref{eq:121}) thus implies
$a^{3}=\frac{\left(-E\right)^{-3}G^{3}\left(m_{1}m_{2}\right)^{3}}{8}$,
and substituting this in (\ref{eq:period-2-1}) gives
\begin{equation}
T^{2}=\frac{\pi^{2}}{2}\left(-E\right)^{-3}G^{2}\left(m_{1}m_{2}\right)^{3}\left(m_{1}+m_{2}\right)^{-1}.\label{eq:period-2-2}
\end{equation}

Unfortunately, the equations of motion (\ref{eq:motion}) do not admit
closed solutions for $n>2$. Even the Newtonian three-body problem
is much more difficult than the two-body problem. While the latter
was essentially solved by Newton \cite{NEWT}%
\begin{comment}
and Johann Bernoulli 
\end{comment}
, the former is a famous problem that has remained recalcitrant despite
efforts by Newton himself, Euler \cite{EUL}, Lagrange \cite{LAG},
Poincaré \cite{POIN} and many others; it is still an active area
of research. Powerful computers have made it practicable to generate
numerical solutions of selected equations of motion, but the theoretical
understanding of the problem is still a work-in-progress. 

In 2018, Bo-Hua Sun \cite{SUN1} proposed a bold conjecture which
relates to Kepler's third law, namely a generalization of (\ref{eq:period-2-2}).
He suggested that the orbital period of a three-body system is given
by the formula
\begin{equation}
T^{2}=\frac{\pi^{2}}{2}\left(-E\right)^{-3}G^{2}\left(\left(m_{1}m_{2}\right)^{3}+\left(m_{1}m_{3}\right)^{3}+\left(m_{2}m_{3}\right)^{3}\right)\left(m_{1}+m_{2}+m_{3}\right)^{-1},\label{eq:period-3}
\end{equation}
and, more generally, that the period of an orbital $n$-body system
is given by\footnote{There is an inaccuracy in \cite{SUN1}, where the double sum is written
as $\sum_{i=1}^{n}\sum_{j=i+1}^{n}\left(m_{i}m_{j}\right)^{3}$.}
\begin{equation}
T^{2}=\phi_{T}\left(m_{1},\ldots,m_{n},E,G\right)=\frac{\pi^{2}}{2}\left(-E\right)^{-3}G^{2}\sum_{i=1}^{n-1}\sum_{j=i+1}^{n}\left(m_{i}m_{j}\right)^{3}\left(\sum_{i=1}^{n}m_{i}\right)^{-1}.\label{eq:period-4-1}
\end{equation}
In addition to the analogy between (\ref{eq:period-2-2}) and (\ref{eq:period-4-1}),
this conjecture is supported by%
\begin{comment}
an informal dimensional argument as well as
\end{comment}
{} known numerical solutions \cite{LI1,LI2} of equations of motion
describing certain three-body systems.

Relating (\ref{eq:period-4-1}) to symmetries of an $n$-body system,
let $\sigma$ be a permutation of $1,\ldots,n$ and $\overline{\sigma}:p_{i}\mapsto p_{\sigma\left(i\right)}$
a corresponding relabeling of $p_{1},\ldots,p_{n}$. Each $\overline{\sigma}$
generates permutations $m_{i}\mapsto m_{\sigma\left(i\right)}$ and
$\mathbf{q}_{i}\mapsto\mathbf{q}_{\sigma\left(i\right)}$ of the $n$-tuples
of scalars $\left(m_{1},\ldots,m_{n}\right)$ and vector-valued functions
$\left(\mathbf{q}_{1},\ldots,\mathbf{q}_{n}\right)$, respectively.
It is shown in Section \ref{subsec:21} that $\overline{\sigma}$
does not affect the system of equations of motion and their initial
conditions, and that therefore $\phi_{T}$, defined by (\ref{eq:period-4-1}),
must satisfy 
\begin{gather}
\phi_{T}\left(m_{1},\ldots,m_{n},E,G\right)=\phi_{T}\left(m_{\sigma\left(1\right)},\ldots,m_{\sigma\left(n\right)},E,G\right)\label{eq:syma}
\end{gather}
for any permutation $\sigma$ of $1,\ldots,n$, as it indeed does.
A deeper question concerns the uniqueness of those functions $\phi_{T}$
that satisfy the symmetries in (\ref{eq:syma}). To the extent that
$\phi_{T}$ is shown to be unique by dimensional analysis, there exists
theoretical support from this point of view for Sun's conjecture.

The main goal of this article is to investigate the uniqueness of
$\phi_{T}$, using augmented dimensional analysis \cite{JON1}, a
generalization of dimensional analysis that, in contrast to classical
dimensional analysis, takes symmetries of the kind considered here
into account. As a consequence, information about $T$ can be extracted
from the parameters of an $n$-body system without actually solving
it, evading the obstacle that such solutions are typically not known
for $n>2$.

Augmented dimensional analysis is surveyed in Section \ref{sec:2},
where we also discuss the symmetry assumptions used. In Sections 3,
4 and 5, we apply dimensional analysis to two-body systems, three-body-systems
and $n$-body systems, respectively, providing support for uniqueness.
In Section 6, an assumption used in generalizations to the $n$-body
case is discussed, and in Section 7 we derive the values of coefficients
of proportionality. In Section 8, four main generalizations to the
$n$-body case are highlighted and compared with results obtained
by Sun \cite{SUN1} and Semay \cite{SEM1}. Section 9 contains final
comments on the uniqueness of the equations derived.

\section{\label{sec:2}Preliminaries}

\subsection{Augmented dimensional analysis}

\selectlanguage{british}%
Dimensional analysis can be described as a method of transforming
''physically meaningful'' equations of the form
\[
t=\phi\left(t_{1},\ldots,t_{n}\right),
\]
describing relationships among quantities, into equivalent but more
informative equations, using data about the dimensions of $t,t_{1},\ldots,t_{n}$.
This can be done by representing the unknown function $\phi$ as a
product of the form $\prod\nolimits_{j=1}^{r}\!x{}_{j}^{\mathfrak{K}_{j}}\psi$,
where $\psi$, while also unspecified, is a function of fewer arguments
than $\phi$ (possibly none, in which case $\psi$ is an unknown constant).

Specifically, for a suitable partition $\left\{ \left\{ y_{1},\ldots,y_{n-r}\right\} ,\left\{ x_{1},\ldots,x_{r}\right\} \right\} $
of $\left\{ t_{1},\ldots,t_{n}\right\} $,
\begin{equation}
t^{\mathfrak{K}}=\prod\nolimits_{j=1}^{r}\!x{}_{j}^{\mathfrak{K}_{j}}\,\psi\left(\pi_{1},\ldots,\pi_{n-r}\right),\label{eq:t^k-1}
\end{equation}
where $\mathfrak{K}$ is a positive integer, $\mathfrak{K}_{j}$ are
integers, $\pi_{1},\ldots,\pi_{n-r}$ and $\psi\left(\pi_{1},\ldots,\pi_{n-r}\right)$
are dimensionless quantities and $\pi_{i}$ is a product of the form
\begin{equation}
\pi_{i}=y_{i}^{k_{i}}\left(\prod\nolimits_{j=1}^{r}\!x{}_{j}^{k_{ij}}\right)^{-1}\quad(i=1,\ldots,n-r),\label{eq:y*_i-1}
\end{equation}
where $k_{i}$ is a positive integer and $k_{ij}$ are integers.

Integers $\mathfrak{K}$, $\mathfrak{K}_{j}$, $k_{i}$ and $k_{ij}$
satisfying (\foreignlanguage{english}{\ref{eq:t^k-1}) and (\ref{eq:y*_i-1})
exist if and only if $\left\{ x_{1},\ldots,x_{r}\right\} $, commonly
called a set of ``repeating variables'', is a set of dimensionally
independent variables which is maximal in $\left\{ t,t_{1},\ldots,t_{n}\right\} $,
so that
\begin{equation}
\left[t\right]^{\mathfrak{K}}=\prod\nolimits_{j=1}^{r}\!\left[x_{j}\right]^{\mathfrak{K}_{j}},\qquad\left[y_{i}\right]^{k_{i}}=\prod\nolimits_{j=1}^{r}\left[x_{j}\right]^{k_{ij}}\quad(i=1,\ldots,n-r),\label{eq:dimexp}
\end{equation}
where $\left[x\right]$ is the dimension of $x$, has solutions }$\mathfrak{K}$,
$\mathfrak{K}_{j}$, $k_{i}$ and $k_{ij}$\foreignlanguage{english}{,
which are unique if we also require that $\mathrm{lcd}\left(\mathfrak{K},\mathfrak{K}_{1},\ldots,\mathfrak{K}_{r}\right)=\mathrm{lcd}\left(k_{i},k_{i1},\ldots,k_{ir}\right)=1$
for $i=1,\ldots,n-r.$}

\selectlanguage{english}%
The expansions of the form (\ref{eq:dimexp}) can be obtained from
dimensional matrices, where each quantity variable corresponds to
a column vector giving the dimen\-sional formula of that quantity
relative to given base dimensions, in mechanics usually $\mathsf{L},\mathsf{T},\mathsf{M}$.
For example, if the dimensional matrix is
\[
\begin{array}{ccccc}
 & \left[t\right] & \left[y_{1}\right] & \left[x_{1}\right] & \left[x_{2}\right]\\
\mathsf{L} & 1 & 0 & 1 & 0\\
\mathsf{T} & -2 & 0 & 0 & 1\\
\mathsf{M} & 1 & 0 & 1 & 0
\end{array},
\]
then the dimensional formulas are $\left[t\right]=\mathsf{L}\mathsf{T}^{-2}\mathsf{M}$,
$\left[y_{1}\right]=\mathsf{L}^{0}\mathsf{T}^{0}\mathsf{M}^{0}$,
$\left[x_{1}\right]=\mathsf{LM}$ and $\left[x_{2}\right]=\mathsf{T}$.
Thus, $\left\{ x_{1},x_{2}\right\} $ is the maximal subset of $\left\{ y_{1},x_{1},x_{2}\right\} $
with independent variables, and we have $\left[t\right]=\left[x_{1}\right]\left[x_{2}\right]^{-2}$
and $\left[y_{1}\right]=\left[x_{1}\right]^{0}\left[x_{2}\right]^{0}$,
so $\left(\mathfrak{K},\mathfrak{K}_{1},\mathfrak{K}_{2}\right)=\left(1,1,-2\right)$
and $\left(k_{1},k_{11},k_{12}\right)=\left(1,0,0\right)$, so (\ref{eq:t^k-1})
becomes $t=x_{1}x_{2}^{-2}\psi\left(y_{1}\right)$.

It is important to note that there may be more than one set of repeating
variables. In augmented dimensional analysis the result obtained is
not one equation of the form (\ref{eq:t^k-1}) but a system of $p\geq1$
such equations,
\begin{equation}
\begin{cases}
t^{\mathfrak{K}_{1}}=\prod\nolimits_{j=1}^{r}x{}_{1j}^{k_{1j}}\,\psi_{1}\left(\pi_{11},\ldots,\pi_{1\left(n-r\right)}\right),\\
\cdots\\
t^{\mathfrak{K}_{p}}=\prod\nolimits_{j=1}^{r}x{}_{pj}^{k_{pj}}\,\psi_{p}\left(\pi_{p1},\ldots,\pi_{p\left(n-r\right)}\right),
\end{cases}\label{eq:dasyst}
\end{equation}
each equation corresponding to a distinct maximal set of dimensionally
independent variables $\left\{ x_{i1},\ldots,x_{ir}\right\} $ \cite{JON1}.
(All $\mathfrak{K}_{i}$ will be equal in the applications below.)
For example, if we had $\left[y_{1}\right]=\mathsf{L}^{0}\mathsf{T^{-1}}\mathsf{M}^{0}$
in the example above then $\left\{ y_{1},x_{1}\right\} $ and $\left\{ x_{1},x_{2}\right\} $
would both be sets of repeating variables, and we would obtain
\[
\begin{cases}
t=y_{1}^{2}x_{1}\,\psi_{1}\left(x_{2}y_{1}\right),\\
t=x_{1}x_{2}^{-2}\,\psi_{2}\left(y_{1}x_{2}\right).
\end{cases}
\]

By combining a system of two or more equations with additional information,
we can derive results that could not have been obtained from these
equations individually, even in combination with the same additional
information. More specifically, while classical dimensional analysis
produces one equation of the form (\ref{eq:t^k-1}), where $\psi$
is unknown, augmented dimensional analysis not only produces a system
of equations (\ref{eq:dasyst}), but may also make it possible, introducing
a symmetry assumption, to find $\psi_{1},\ldots,\psi_{p}$, and hence
also $\phi$, up to a multiplicative constant. This principle, first
formulated and demonstrated in \cite{JON2}, will be used repeatedly
in this article. 

\subsection{\label{subsec:21}Symmetries}

An $n$-body system has important symmetries which are encoded in
its equations of motion and their initial conditions. Consider the
system of equations
\begin{equation}
\begin{cases}
m_{1}\ddot{\mathbf{q}}_{1}=G\sum_{j=2}^{n}m_{1}m_{j}\frac{\left(\mathbf{q}_{j}-\mathbf{q}_{1}\right)}{\left\Vert \mathbf{q}_{j}-\mathbf{q}_{1}\right\Vert ^{3}},\quad\mathbf{q}_{1}\left(0\right)=\mathrm{q}_{1},\quad\dot{\mathbf{q}}{}_{1}\left(0\right)=\dot{\mathrm{q}}_{1} & \left(1\right),\\
\cdots\\
m_{n}\ddot{\mathbf{q}}_{n}=G\sum_{j=1}^{n-1}m_{n}m_{j}\frac{\left(\mathbf{q}_{j}-\mathbf{q}_{n}\right)}{\left\Vert \mathbf{q}_{j}-\mathbf{q}_{n}\right\Vert ^{3}},\quad\mathbf{q}_{n}\left(0\right)=\mathrm{q}_{1},\quad\dot{\mathbf{q}}_{n}\left(0\right)=\dot{\mathrm{q}}_{n} & \left(n\right),
\end{cases}\label{eq:eom-init}
\end{equation}
which contains all equations in (\ref{eq:motion}) together with their
initial conditions. Let again $\overline{\sigma}:p_{i}\mapsto p_{\sigma\left(i\right)}$
be a relabeling of the bodies in the $n$-body system $p_{1},\ldots,p_{n}$.
As $m_{i}$ is the mass of $p_{i}$ and $\mathbf{q}_{i}\left(t\right)$
the radius vector of $p_{i}$ at $t$, $\overline{\sigma}$ corresponds
to permutations $m_{i}\mapsto m_{\sigma\left(i\right)}$ and $\mathbf{q}_{i}\mapsto\mathbf{q}_{\sigma\left(i\right)}$
of the n-tuples of scalars $\left(m_{1},\ldots,m_{n}\right)$ and
vector-valued functions $\left(\mathbf{q}_{1},\ldots,\mathbf{q}_{n}\right)$,
respectively, so $\overline{\sigma}$ sends the row
\[
m_{i}\ddot{\mathbf{q}}_{i}=G\sum_{j\in\mathbf{n}\setminus\left\{ i\right\} }m_{i}m_{j}\frac{\left(\mathbf{q}_{j}-\mathbf{q}_{i}\right)}{\left\Vert \mathbf{q}_{j}-\mathbf{q}_{i}\right\Vert ^{3}},\quad\mathbf{q}_{i}\left(0\right)=\mathrm{q}_{i},\quad\dot{\mathbf{q}}_{i}\left(0\right)=\dot{\mathrm{q}}_{i}\qquad\left(i\right)
\]
in (\ref{eq:eom-init}) to the row
\begin{align*}
m_{i'}\ddot{\mathbf{q}}_{i'}=G\sum_{j'\in\mathbf{n}\setminus\left\{ i'\right\} }m_{i'}m_{j'}\frac{\left(\mathbf{q}_{j'}-\mathbf{q}_{i'}\right)}{\left\Vert \mathbf{q}_{j'}-\mathbf{q}_{i'}\right\Vert ^{3}},\quad\mathbf{q}_{i'}\left(0\right)=\mathrm{q}_{i'},\quad\dot{\mathbf{q}}_{i'}\left(0\right)=\dot{\mathrm{q}}_{i'} & \qquad\left(i'\right),
\end{align*}
where $i'=\sigma\left(i\right)$ and $j'=\sigma\left(j\right)$. Thus,
$\overline{\sigma}$ just permutes the rows in (\ref{eq:eom-init}),
so it is a mapping of (\ref{eq:eom-init}) to itself, since two equation
systems with the same set of differential equations and corresponding
initial conditions are equal regardless of the order in which the
equations and conditions are listed. Hence, any quantity value determined
by (\ref{eq:eom-init}) is invariant under a relabeling $\overline{\sigma}$
of $p_{1},\ldots,p_{n}$.

Consider the three functions 
\begin{gather*}
\phi_{T}:\left(m_{1},\ldots,m_{n},E,G\right)\mapsto T^{\mathfrak{J}},\qquad\varphi{}_{T}:\left(m_{1},\ldots,m_{n},d,G\right)\mapsto T^{\mathfrak{K}},\\
\phi_{d}:\left(m_{1},\ldots,m_{n},E,G\right)\mapsto d^{\mathfrak{L}}.
\end{gather*}
The distance $d$ is assumed to be given by a function $f$ such that
\[
d=f\left(\mathbf{q}_{1},\ldots,\mathbf{q}_{n}\right)=f\left(\mathbf{q}_{\sigma\left(1\right)},\ldots,\mathbf{q}_{\sigma\left(n\right)}\right)
\]
for every permutation $\sigma$ of $1,\ldots,n$; we say that $d$
is a \emph{symmetric distance}. Also, the total mechanical energy
$E$ is given by
\begin{align*}
E & =\sum_{i=1}^{n}\frac{m_{i}^{2}}{2}\left(\mathrm{\dot{q}}_{i}\right)^{2}-G\sum_{i=1}^{n-1}\sum_{j=i+1}^{n}\frac{m_{i}m_{j}}{\left\Vert \mathrm{q}_{j}-\mathrm{q}_{i}\right\Vert }\\
 & =\sum_{i'=1}^{n}\frac{m_{i'}^{2}}{2}\left(\mathrm{\dot{q}}_{i'}\right)^{2}-G\sum_{i'=1}^{n-1}\sum_{j'=i+1}^{n}\frac{m_{i'}m_{j'}}{\left\Vert \mathrm{q}_{j'}-\mathrm{q}_{i'}\right\Vert },
\end{align*}
where $i'=\sigma\left(i\right)$ and $j'=\sigma\left(j\right)$. Thus,
$d$ and $E$ by definition do not depend on $\overline{\sigma}$,
so as $T$ is determined by (\ref{eq:eom-init}) regardless of $\overline{\sigma}$
we have
\begin{gather}
T^{\mathfrak{J}}=\phi_{T}\left(m_{1},\ldots,m_{n},E,G\right)=\phi_{T}\left(m_{\sigma\left(1\right)},\ldots,m_{\sigma\left(n\right)},E,G\right),\label{eq:2ym1}\\
T^{\mathfrak{K}}=\varphi_{T}\left(m_{1},\ldots,m_{n},d,G\right)=\varphi_{T}\left(m_{\sigma\left(1\right)},\ldots,m_{\sigma\left(n\right)},d,G\right),\label{eq:sym3}\\
d^{\mathfrak{L}}=\phi_{d}\left(m_{1},\ldots,m_{n},E,G\right)=\phi_{d}\left(m_{\sigma\left(1\right)},\ldots,m_{\sigma\left(n\right)},E,G\right)\label{eq:sym2}
\end{gather}
for any $\phi_{T},\varphi{}_{T},\phi_{d}$ and any permutation $\sigma$
of $1,\ldots,n$.

\section{\label{sec:3}Two-body systems}

In this section, we give examples of augmented dimensional analysis
related to equations (\ref{eq:period-2-1}) and (\ref{eq:period-2-2})
above.
\selectlanguage{british}%
\begin{example}
\label{ex1}\foreignlanguage{english}{$T^{\mathfrak{K}}=\varphi{}_{T}\left(m_{1},m_{2},d,G\right)$}

Let two bodies with masses $m_{1}$ and $m_{2}$ revolve around each
other in empty space under influence of their mutual gravitational
attraction. Assume that the orbital period $T$ depends on the two
masses, a symmetric characteristic distance $d$ and the gravitational
constant $G$, that is, $T^{\mathfrak{K}}=\phi_{T}\left(m_{1},m_{2},d,G\right)$.
\foreignlanguage{english}{The dimensional matrix expressing the dimensionality
of the variables in this equation relative to the base dimensions
$\mathsf{L},\mathsf{T},\mathsf{M}$ is}

\selectlanguage{english}%
\[
\begin{array}{cccccc}
 & \left[T\right] & \left[m_{1}\right] & \left[m_{2}\right] & \left[d\right] & \left[G\right]\\
\mathsf{L} & 0 & 0 & 0 & 1 & 3\\
\mathsf{T} & 1 & 0 & 0 & 0 & -2\\
\mathsf{M} & 0 & 1 & 1 & 0 & -1
\end{array}.
\]
 The sets of repeating variables are $\left\{ m_{1},d,G\right\} $
and $\left\{ m_{2},d,G\right\} $, and the corresponding system of
equations, where $\mathfrak{K}=2$, is
\begin{equation}
\begin{cases}
T^{2}=m_{1}^{-1}d^{3}G^{-1}\,\psi_{1}\left(m_{2}/m_{1}\right), & (A1)\\
T^{2}=m_{2}^{-1}d^{3}G^{-1}\,\psi_{2}\left(m_{1}/m_{2}\right). & (A2)
\end{cases}\label{eq:sys11}
\end{equation}

If $\varphi_{T}\left(m_{1},m_{2},d,G\right)=\varphi_{T}\left(m_{2},m_{1},d,G\right)$,
in accordance with (\ref{eq:sym3}), then exchanging $m_{1}$ and
$m_{2}$ in ($A1$) gives
\[
T^{2}=m_{2}^{-1}d^{3}G^{-1}\,\psi_{1}\left(m_{1}/m_{2}\right)\qquad(A1'),
\]
and it follows from $(A1')$ and (A2) that $\psi_{1}=\psi_{2}=\psi$.
Thus, (\ref{eq:sys11}) implies
\[
m_{1}^{-1}\psi\left(m_{2}/m_{1}\right)=m_{2}^{-1}\psi\left(m_{1}/m_{2}\right),
\]
 so setting $x=m_{2}/m_{1}$ we obtain the functional equation \foreignlanguage{british}{
\begin{equation}
\psi\left(x\right)=x^{-1}\,\psi\left(x^{-1}\right).\label{eq:fe1}
\end{equation}
}

\selectlanguage{british}%
(\foreignlanguage{english}{\ref{eq:fe1})} has solutions of the form
\begin{equation}
\psi(x)=c_{T}\left(1+x\right)^{-1}\label{eq:sol1}
\end{equation}
since $\frac{1}{x}c_{T}\frac{1}{1+x^{-1}}=c_{T}\frac{1}{x+1}$, and
substituting (\foreignlanguage{english}{\ref{eq:sol1})} in ($A1$)
gives
\begin{equation}
T^{2}=c_{T}d^{3}G^{-1}(m_{1}+m_{2})^{-1}\label{eq:33}
\end{equation}
since $\frac{1}{m_{1}}\frac{1}{1+m_{2}/m_{1}}=\frac{1}{m_{1}+m_{2}}$.

\smallskip{}

\begin{comment}
(\foreignlanguage{english}{\ref{eq:33}) can also be derived from
(\ref{eq:sol1}) and ($A2$) by setting $x=m_{1}/m_{2}$. }
\end{comment}
\foreignlanguage{english}{Note that the semi-major axis $a$ is a
symmetric distance by definition}%
\begin{comment}
\selectlanguage{english}%
and means are symmetric
\end{comment}
\foreignlanguage{english}{ and a characteristic distance in view of
(\ref{eq:period-2-1}), so we can recover (\ref{eq:period-2-1}) from
(\ref{eq:33}) by setting $d=a$ and $c_{T}=4\pi^{2}.$}
\end{example}

\selectlanguage{english}%
\selectlanguage{british}%
\begin{example}
$T^{2}=c_{T}d^{3}G^{-1}(m_{1}+m_{2})^{-1}$, $d^{\mathfrak{L}}=\phi_{d}\left(m_{1},m_{2},E,G\right)$

\selectlanguage{english}%
\label{ex2}To replace $d$ by $E$ in (\ref{eq:period-2-1}), we
express $d$ in terms of $m_{1}$, $m_{2}$, $E$ and $G$. That is,
we seek to find a function $\phi_{d}$ such that $d^{\mathfrak{L}}=\phi_{d}\left(m_{1},m_{2},E,G\right)$.
The dimensional matrix is

\[
\begin{array}{cccccc}
 & \left[d\right] & \left[m_{1}\right] & \left[m_{2}\right] & \left[E\right] & \left[G\right]\\
\mathsf{L} & 1 & 0 & 0 & 2 & 3\\
\mathsf{T} & 0 & 0 & 0 & -2 & -2\\
\mathsf{M} & 0 & 1 & 1 & 1 & -1
\end{array}.
\]
 The sets of repeating variables are $\left\{ m_{1},E,G\right\} $
and $\left\{ m_{2},E,G\right\} $, and the corresponding system of
equations, where $\mathfrak{K}=1$, is
\begin{equation}
\begin{cases}
d=m_{1}^{2}\left(-E\right)^{-1}G\,\psi_{1}\left(m_{2}/m_{1}\right), & (B1)\\
d=m_{2}^{2}\left(-E\right)^{-1}G\,\psi_{2}\left(m_{1}/m_{2}\right). & (B2)
\end{cases}\label{eq:12-1}
\end{equation}
By (\ref{eq:sym2}), exchanging $m_{1}$ and $m_{2}$ in $(B1)$ gives
\[
d=m_{2}^{2}\left(-E\right)^{-1}G\,\psi_{1}\left(m_{1}/m_{2}\right)\qquad(B1'),
\]
and it follows from $(B1')$ and (B2) that $\psi_{1}=\psi_{2}=\psi$.
Thus, (\ref{eq:12-1}) implies
\[
m_{1}^{2}\left(-E\right)^{-1}G\,\psi\left(m_{2}/m_{1}\right)=m_{2}^{2}\left(-E\right)^{-1}G\,\psi\left(m_{1}/m_{2}\right),
\]
so setting $x=m_{2}/m_{1}$ we have the functional equation
\begin{equation}
\psi\left(x\right)=x^{2}\psi\left(x^{-1}\right).\label{eq:fun2}
\end{equation}
(\ref{eq:t-3}) has solutions of the form
\begin{equation}
\psi\left(x\right)=Cx\label{eq:solutionb2-1-1}
\end{equation}
since $x^{2}Cx^{-1}=Cx$.%
\begin{comment}
\selectlanguage{british}%
(\foreignlanguage{english}{\ref{eq:formul2-1}) can also be derived
from (\ref{eq:solutionb2-1-1}) and ($B2$) by setting $x=m_{1}/m_{2}$}
\end{comment}

Substituting (\ref{eq:solutionb2-1-1}) in $(B1)$, we obtain
\begin{equation}
d=C\left(-E\right)G^{1}\left(m_{1}m_{2}\right)\label{eq:formul2-1}
\end{equation}
since $m_{1}^{2}\left(m_{2}/m_{1}\right)=m_{1}m_{2}.$

Finally, setting $k_{d}=C^{3}$ and combining (\ref{eq:33}) and (\ref{eq:formul2-1})
gives
\begin{equation}
T^{2}=c_{T}k_{d}\left(-E\right)^{-3}G^{2}\left(m_{1}m_{2}\right)^{3}\left(m_{1}+m_{2}\right)^{-1},\label{eq:T-E-2-1}
\end{equation}
and setting $C=1/2$ so that $c_{T}k_{d}=\left(4\pi^{2}\right)/8=\pi^{2}/2$
we recover (\ref{eq:period-2-2}).
\end{example}

\selectlanguage{english}%

\section{\label{sec:4}Three-body systems}

When going from $n=2$ to $n=3$, equations change form, and the three-body
case may be said to be qualitatively more similar to the $n$-body
case than to the two-body case. It is therefore useful to spell out
the details in the three-body case before proceeding to the general
case. %
\begin{comment}
While all orbital trajectories of two-body systems have the same elliptic
shape, a complication for $n>2$ is that orbital trajectories may
have quite different shapes. As a consequence, the constants $4\pi^{2}$
and $\pi^{2}/2$, which work for all elliptic orbits, may have different
values for orbital trajectories of different shapes. Also, $a$ cannot
be defined as the mean extremal distance between the two bodies for
$n>2$, but we assume here that, as in the two-body case, there exists
a characteristic distance $d$ which is invariant under a permutation
of the positions of the three bodies.
\end{comment}

\begin{example}
$T^{\mathfrak{K}}=\varphi{}_{T}\left(m_{1},m_{2},m_{3},d,G\right)$

Assume that the orbital period of a three-body system $T$ is determined
by the masses $m_{1},m_{2},m_{3}$ of the three bodies, a symmetric
characteristic distance $d$, and the gravitational constant $G$;
that is, $T^{\mathfrak{K}}=\phi_{T}\left(m_{1},m_{2},m_{3},d,G\right)$.
The dimensional matrix corresponding to this equation is
\[
\begin{array}{ccccccc}
 & \left[T\right] & \left[m_{1}\right] & \left[m_{2}\right] & \left[m_{3}\right] & \left[d\right] & \left[G\right]\\
\mathsf{L} & 0 & 0 & 0 & 0 & 1 & 3\\
\mathsf{T} & 1 & 0 & 0 & 0 & 0 & -2\\
\mathsf{M} & 0 & 1 & 1 & 1 & 0 & -1
\end{array},
\]
and the sets of repeating variables are $\left\{ m_{1},d,G\right\} $,
$\left\{ m_{2},d,G\right\} $ and $\left\{ m_{3},d,G\right\} $. Corresponding
to the six equations of the form 
\[
T^{\mathfrak{K}}=\varphi_{T}\left(m_{\text{\ensuremath{\sigma}\ensuremath{\left(1\right)}}},m_{\sigma\left(2\right)},m_{\sigma\left(3\right)},d,G\right),
\]
where $\sigma$ is a permutation of $1,2,3$, we obtain the equation
system
\begin{equation}
\begin{cases}
T^{2}=m_{1}^{-1}d^{3}G^{-1}\,\psi_{11}\left(m_{2}/m_{1},m_{3}/m_{1}\right),\\
T^{2}=m_{1}^{-1}d^{3}G^{-1}\,\psi_{12}\left(m_{3}/m_{1},m_{2}/m_{1}\right),\\
T^{2}=m_{2}^{-1}d^{3}G^{-1}\,\psi_{21}\left(m_{1}/m_{2},m_{3}/m_{2}\right),\\
T^{2}=m_{2}^{-1}d^{3}G^{-1}\,\psi_{22}\left(m_{3}/m_{2},m_{1}/m_{2}\right),\\
T^{2}=m_{3}^{-1}d^{3}G^{-1}\,\psi_{31}\left(m_{1}/m_{3},m_{2}/m_{3}\right),\\
T^{2}=m_{3}^{-1}d^{3}G^{-1}\,\psi_{32}\left(m_{2}/m_{3},m_{1}/m_{3}\right).
\end{cases}\label{eq:20}
\end{equation}

We have $\psi_{11}\left(x,y\right)=\psi_{12}\left(y,x\right)$, $\psi_{21}\left(x,y\right)=\psi_{22}\left(y,x\right)$
and $\psi_{31}\left(x,y\right)=\psi_{32}\left(y,x\right)$, so we
can reduce (\ref{eq:20}) without loss of information to
\begin{equation}
\begin{cases}
T^{2}=m_{1}^{-1}d^{3}G^{-1}\,\psi_{1}\left(m_{2}/m_{1},m_{3}/m_{1}\right), & (A1)\\
T^{2}=m_{2}^{-1}d^{3}G^{-1}\,\psi_{2}\left(m_{1}/m_{2},m_{3}/m_{2}\right), & (A2)\\
T^{2}=m_{3}^{-1}d^{3}G^{-1}\,\psi_{3}\left(m_{1}/m_{3},m_{2}/m_{3}\right), & (A3)
\end{cases}\label{eq:21}
\end{equation}
where $\psi_{1}=\psi_{11}$, $\psi_{2}=\psi_{21}$ and $\psi_{3}=\psi_{31}$.

In view of (\ref{eq:sym3}), exchanging $m_{1}$ and $m_{2}$ in ($A1$)
gives
\[
T^{2}=m_{2}^{-1}d^{3}G^{-1}\,\psi_{1}\left(m_{1}/m_{2},m_{3}/m_{2}\right)\quad(A1'),
\]
and it follows from $(A1')$ and (A2) that $\psi_{1}=\psi_{2}=\psi$.
\begin{comment}
We can show similarly that $\psi_{2}=\psi_{3}$, so $\psi_{1}=\psi_{2}=\psi_{3}=\psi$\foreignlanguage{british}{. }
\end{comment}
\foreignlanguage{british}{Combining ($A1$) and ($A2$) we obtain
\[
m_{1}^{-1}d^{3}G^{-1}\,\psi\left(m_{2}/m_{1},m_{3}/m_{1}\right)=m_{2}^{-1}d^{3}G^{-1}\,\psi\left(m_{1}/m_{2},m_{3}/m_{2}\right),
\]
so setting $x=m_{2}/m_{1}$ and $y=m_{3}/m_{1},$ so that $x^{-1}y=m_{3}/m_{2}$,
we obtain the functional equation 
\begin{equation}
\psi\left(x,y\right)=x^{-1}\psi\left(x^{-1},x^{-1}y\right).\label{eq:fe4}
\end{equation}
}%
\begin{comment}
\selectlanguage{british}%
and we obtain the same functional equation for each of the two other
pairs of equations in (\foreignlanguage{english}{\ref{eq:21}}) by
choosing $x$ and $y$ appropriately.
\end{comment}

\selectlanguage{british}%
(\foreignlanguage{english}{\ref{eq:fe4})} has solutions 
\begin{equation}
\psi\left(x,y\right)=c_{T}\left(1+x+y\right)^{-1}\label{eq:treb}
\end{equation}
 since $\frac{1}{x}c_{T}\frac{1}{1+x^{-1}+x^{-1}y}=c_{T}\frac{1}{x+1+y}$,
Thus, substituting (\foreignlanguage{english}{\ref{eq:treb}}) in
($A1$) we obtain
\begin{equation}
T^{2}=c_{T}d^{3}G^{-1}\left(m_{1}+m_{2}+m_{3}\right)^{-1}\label{eq:t-3}
\end{equation}
since $\frac{1}{m_{1}}\frac{1}{1+m_{2}/m_{1}+m_{3}/m_{1}}=\frac{1}{m_{1}+m_{2}+m_{3}}$.
\end{example}

\selectlanguage{british}%
\selectlanguage{english}%
\begin{example}
$T^{2}=c_{T}d^{3}G^{-1}(m_{1}+m_{2}+m_{3})^{-1}$, $d^{\mathfrak{L}}=\phi_{d}\left(m_{1},m_{2},m_{3},E,G\right)$

Assuming that $d$ depends on $E$ through the relation $d^{\mathfrak{L}}=\phi_{d}\left(m_{1},m_{2},m_{3},E,G\right)$,
we now use dimensional analysis again to find $\phi_{d}.$ The corresponding
dimensional matrix is

\[
\begin{array}{ccccccc}
 & \left[d\right] & \left[m_{1}\right] & \left[m_{2}\right] & \left[m_{3}\right] & \left[E\right] & \left[G\right]\\
\mathsf{L} & 1 & 0 & 0 & 0 & 2 & 3\\
\mathsf{T} & 0 & 0 & 0 & 0 & -2 & -2\\
\mathsf{M} & 0 & 1 & 1 & 1 & 1 & -1
\end{array},
\]
the sets of repeating variables are $\left\{ m_{1},E,G\right\} $,
$\left\{ m_{2},E,G\right\} $ and $\left\{ m_{3},E,G\right\} $, and
the reduced equation system, obtained as described above, is
\begin{equation}
\begin{cases}
d=m_{1}^{2}\left(-E\right)^{-1}G\,\psi_{1}\left(m_{2}/m_{1},m_{3}/m_{1}\right), & (B1)\\
d=m_{2}^{2}\left(-E\right)^{-1}G\,\psi_{2}\left(m_{1}/m_{2},m_{3}/m_{2}\right), & (B2)\\
d=m_{3}^{2}\left(-E\right)^{-1}G\,\psi_{3}\left(m_{1}/m_{3},m_{2}/m_{3}\right). & (B3)
\end{cases}\label{eq:22-1}
\end{equation}

By (\ref{eq:sym2}), exchanging $m_{1}$ and $m_{2}$ in ($B1$) gives
\[
d=m_{2}^{2}\left(-E\right)^{-1}G\,\psi_{1}\left(m_{1}/m_{2},m_{3}/m_{2}\right)\quad(B1'),
\]
and it follows from $(B1')$ and (B2) that $\psi_{1}=\psi_{2}=\psi$,
so from (B1) and (B2) we obtain
\[
m_{1}^{2}\left(-E\right)^{-1}G\,\psi\left(m_{2}/m_{1},m_{3}/m_{1}\right)=m_{2}^{2}\left(-E\right)^{-1}G\,\psi\left(m_{1}/m_{2},m_{3}/m_{2}\right),
\]
and we can again derive a functional equation, in this case
\begin{equation}
\psi\left(x,y\right)=x^{2}\psi\left(x^{-1},x^{-1}y\right).\label{eq:fe5}
\end{equation}
where $x=m_{2}/m_{1}$ and $y=m_{3}/m_{1}$ so that $x^{-1}y=m_{3}/m_{2}$.%
\begin{comment}
\selectlanguage{british}%
Defining $x$ and $y$ appropriately, we get the same result for each
of the two other pairs of equations in (\foreignlanguage{english}{\ref{eq:22-1}}). 
\end{comment}
\end{example}

\begin{casenv}
\item (\ref{eq:fe5}) has solutions of the form
\begin{equation}
\psi\left(x,y\right)=C\left(x+y+xy\right)\label{eq:formul3-1}
\end{equation}
since $x^{2}C\left(x^{-1}+x^{-1}y+x^{-1}\left(x^{-1}y\right)\right)=C\left(x+xy+y\right).$
Thus, substituting (\ref{eq:formul3-1}) in $(B1)$ we obtain
\begin{equation}
d=C\left(-E\right)^{-1}G\left(m_{1}m_{2}+m_{1}m_{3}+m_{2}m_{3}\right)\label{eq:d-3-1}
\end{equation}
since $m_{1}^{2}\left(\frac{m_{2}}{m_{1}}+\frac{m_{3}}{m_{1}}+\frac{m_{2}}{m_{1}}\frac{m_{3}}{m_{1}}\right)=m_{1}m_{2}+m_{1}m_{3}+m_{2}m_{3}$.
\begin{comment}
\selectlanguage{british}%
We get the same result for each of the two other equations in (\foreignlanguage{english}{\ref{eq:22}}). 
\end{comment}
Finally, combining (\ref{eq:t-3}) and (\ref{eq:d-3-1}) and setting
$k_{d}^{\left(1\right)}=C^{3}$, we obtain
\[
T^{2}=c_{T}k_{d}^{\left(1\right)}\left(-E\right)^{-3}G^{2}\left(m_{1}m_{2}+m_{1}m_{3}+m_{2}m_{3}\right)^{3}\left(m_{1}+m_{2}+m_{3}\right)^{-1}.
\]
\item (\ref{eq:fe5}) also has solutions of the form
\begin{equation}
\psi\left(x,y\right)=\tilde{C}\sqrt[3]{x^{3}+y^{3}+x^{3}y^{3}}\label{eq:formul3-1-1}
\end{equation}
since
\begin{align*}
\left(x^{2}\psi\left(x^{-1},x^{-1}y\right)\right)^{3} & =\left(x^{2}\tilde{C}\sqrt[3]{x^{-3}+\left(x^{-1}y\right)^{3}+x^{-3}\left(x^{-1}y\right)^{3}}\right)^{3}\\
 & =x^{6}\tilde{C}^{3}\left(x^{-3}+x^{-3}y^{3}+x^{-3}\left(x^{-3}y^{3}\right)\right)=\tilde{C}^{3}\left(x^{3}+x^{3}y^{3}+y^{3}\right)
\end{align*}
Thus, substituting (\ref{eq:formul3-1-1}) in $(B1)$ we obtain
\begin{equation}
d^{3}=\tilde{C}^{3}\left(-E\right)^{-3}G^{3}\left(\left(m_{1}m_{2}\right)^{3}+\left(m_{1}m_{3}\right)^{3}+\left(m_{2}m_{3}\right)^{3}\right)\label{eq:d-3-2}
\end{equation}
since $m_{1}^{6}\left(\frac{m_{2}^{3}}{m_{1}^{3}}+\frac{m_{3}^{3}}{m_{1}^{3}}+\frac{m_{2}^{3}}{m_{1}^{3}}\frac{m_{3}^{3}}{m_{1}^{3}}\right)=m_{1}^{3}m_{2}^{3}+m_{1}^{3}m_{3}^{3}+m_{2}^{3}m_{3}^{3}$.
Finally, combining (\ref{eq:t-3}) and (\ref{eq:d-3-2}) and setting
$k_{d}^{\left(3\right)}=\tilde{C}^{3}$, we obtain
\[
T^{2}=c_{T}k_{d}^{\left(3\right)}\left(-E\right)^{-3}G^{2}\left(\left(m_{1}m_{2}\right)^{3}+\left(m_{1}m_{3}\right)^{3}+\left(m_{2}m_{3}\right)^{3}\right)\left(m_{1}+m_{2}+m_{3}\right)^{-1},
\]
consistent with (\ref{eq:period-3}).
\end{casenv}
\medskip{}

Note that we have derived two formulas for $T^{2}$, both of which
generalize (\ref{eq:period-2-2}). Thus, the symmetry condition in
(\ref{eq:syma}) does not guarantee uniqueness (up to $c_{T}k_{d}^{\left(1\right)}$
or $c_{T}k_{d}^{\left(3\right)}$) of functions obtained by augmented
dimensional analysis in this case.

\section{\label{sec:5}$N$-body systems}

There is a straightforward generalization from the three-body case
to the $n$-body case. Corresponding to $T^{\mathfrak{K}}=\varphi{}_{T}\left(m_{1},\ldots,m_{n},d,G\right)$,
where $d$ is again a symmetric characteristic distance, we have the
dimensional matrix
\[
\begin{array}{ccccccc}
 & \left[T\right] & \left[m_{1}\right] & \ldots & \left[m_{n}\right] & \left[d\right] & \left[G\right]\\
\mathsf{L} & 0 & 0 & \ldots & 0 & 1 & 3\\
\mathsf{T} & 1 & 0 & \ldots & 0 & 0 & -2\\
\mathsf{M} & 0 & 1 &  & 1 & 0 & -1
\end{array},
\]
the sets of repeating variables $\left\{ m_{1},d,G\right\} ,\ldots,\left\{ m_{n},d,G\right\} $,
and the full system of $n!$ equations
\begin{equation}
\begin{cases}
T^{2}=m_{i}^{-1}d^{3}G^{-1}\,\psi_{i\sigma_{i}\left(j_{i1}\right)\cdots\sigma_{i}\left(j_{i\left(n-1\right)}\right)}\left(\frac{m_{\sigma_{i}\left(j_{i1}\right)}}{m_{i}},\ldots,\frac{m_{\sigma_{i}\left(j_{i\left(n-1\right)}\right)}}{m_{i}}\right),\end{cases}\label{eq:310}
\end{equation}
 where $i=1,\ldots,n$, $\left(j_{i1},\ldots,j_{i\left(n-1\right)}\right)=\left(1,\ldots,i-1,i+1,\ldots,n\right)$
and each $\sigma_{i}$ ranges over all permutations of $j_{i1},\ldots,j_{i\left(n-1\right)}$.

Setting 
\[
\psi_{i}\left(m_{j_{i1}}/m_{i},\ldots,m_{j_{i\left(n-1\right)}}/m_{i}\right)=\psi_{ij_{i1}\cdots j_{i\left(n-1\right)}}\left(m_{j_{i1}}/m_{i},\ldots,m_{j_{i\left(n-1\right)}}/m_{i}\right)
\]
 for $i=1,\ldots,n$, we obtain from (\ref{eq:310}) the equivalent
reduced equation system
\begin{equation}
\begin{cases}
T^{2}=m_{1}^{-1}d^{3}G^{-1}\,\psi_{1}\left(m_{2}/m_{1},m_{3}/m_{1},\ldots,m_{n}/m_{1}\right), & (A1)\\
T^{2}=m_{2}^{-1}d^{3}G^{-1}\,\psi_{2}\left(m_{1}/m_{2},m_{3}/m_{2},\ldots,m_{n}/m_{2}\right), & (A2)\\
\cdots\\
T^{2}=m_{n}^{-1}d^{3}G^{-1}\,\psi_{n}\left(m_{1}/m_{n},m_{2}/m_{n},\ldots,m_{n-1}/m_{n}\right). & (An)
\end{cases}\label{eq:31}
\end{equation}
\begin{comment}
where $\psi_{i}\left(m_{j_{i1}}/m_{i},\ldots,m_{j_{i\left(n-1\right)}}/m_{i}\right)=\psi_{ij_{i1}\cdots j_{i\left(n-1\right)}}\left(m_{j_{i1}}/m_{i},\ldots,m_{j_{i\left(n-1\right)}}/m_{i}\right)$,
since 
\begin{gather*}
\psi_{i\sigma_{i}\left(j_{i1}\right)\cdots\sigma_{i}\left(j_{i\left(n-1\right)}\right)}\left(m_{\sigma_{i}\left(j_{i1}\right)}/m_{i},\ldots,m_{\sigma_{i}\left(j_{i\left(n-1\right)}\right)}/m_{i}\right)\\
=\psi_{ij_{i1}\cdots j_{i\left(n-1\right)}}\left(m_{j_{i1}}/m_{i},\ldots,m_{j_{i\left(n-1\right)}}/m_{i}\right)
\end{gather*}
\end{comment}

In view of (\ref{eq:sym3}), exchanging $m_{1}$ and $m_{2}$ in ($A1$)
gives
\[
T^{2}=m_{2}^{-1}d^{3}G^{-1}\,\psi_{1}\left(m_{1}/m_{2},m_{3}/m_{2},\ldots,m_{n}/m_{2}\right)\quad(A1'),
\]
and it follows from $(A1')$ and (A2) that $\psi_{1}=\psi_{2}=\psi$.
\foreignlanguage{british}{Thus, ($A1$) and ($A2$) implies}
\begin{equation}
m_{1}^{-1}\psi\left(m_{2}/m_{1},m_{3}/m_{1},\ldots,m_{n}/m_{1}\right)=m_{2}^{-1}\psi\left(m_{1}/m_{2},m_{3}/m_{2},\ldots,m_{n}/m_{2}\right).\label{eq:301}
\end{equation}
 Setting $x_{i}=m_{i+1}/m_{1}$ for $i=1,\ldots,n-1$, so that $x_{1}^{-1}x_{i}=m_{i+1}/m_{2}$,
(\ref{eq:301}) implies the functional equation
\begin{equation}
\psi\left(x_{1},x_{2},\ldots,x_{n-1}\right)=x_{1}^{-1}\psi\left(x_{1}^{-1},x_{1}^{-1}x_{2},\ldots,x_{1}^{-1}x_{n-1}\right),\label{eq:302}
\end{equation}
with solutions
\begin{equation}
\psi\left(x_{1},x_{2},\ldots,x_{n-1}\right)=c_{T}\left(1+x_{1}+x_{2}\text{+}\ldots+x_{n-1}\right)^{-1}\label{sol31}
\end{equation}
since $\frac{1}{x_{1}}c_{T}\frac{1}{1+x_{1}^{-1}+x_{1}^{-1}x_{2}+\ldots+x_{1}^{-1}x_{n-1}}=c_{T}\frac{1}{x_{1}+1+x_{2}+\ldots+x_{n-1}}$%
\begin{comment}
. Choosing $x_{i}$ appropriately, we obtain the functional equation
(\ref{eq:302}) for each pair of equations in (\ref{eq:31}).
\end{comment}
.

\smallskip{}

\selectlanguage{british}%
Substituting (\foreignlanguage{english}{\ref{sol31}) in} ($A1$),
we obtain\foreignlanguage{english}{
\begin{equation}
T^{2}=c_{T}d^{3}G^{-1}\left(\sum_{i=1}^{n}m_{i}\right)^{-1}\label{eq:T3}
\end{equation}
 since $\frac{1}{m_{1}}\frac{1}{1+m_{2}/m_{1}+m_{3}/m_{1}+\ldots+m_{n}/m_{1}}=\frac{1}{m_{1}+m_{2}+m_{3}+\ldots+m_{n}}$.}

\selectlanguage{english}%
\medskip{}

Also, assuming as before that $d$ depends on $E$ through the relation
\[
d^{\mathfrak{L}}=\phi_{d}\left(m_{1},\ldots,m_{n},E,G\right),
\]
we use dimensional analysis to find $\phi_{d}$. We have the dimensional
matrix
\[
\begin{array}{ccccccc}
 & \left[d\right] & \left[m_{1}\right] & \ldots & \left[m_{3}\right] & \left[E\right] & \left[G\right]\\
\mathsf{L} & 1 & 0 & \ldots & 0 & 2 & 3\\
\mathsf{T} & 0 & 0 & \ldots & 0 & -2 & -2\\
\mathsf{M} & 0 & 1 & \ldots & 1 & 1 & -1
\end{array},
\]
and the sets of repeating variables $\left\{ m_{1},E,G\right\} ,\ldots,\left\{ m_{n},E,G\right\} $.
Reasoning as above, we obtain the reduced equation system
\begin{equation}
\begin{cases}
d=m_{1}^{2}\left(-E\right)^{-1}G\,\psi_{1}\left(m_{2}/m_{1},m_{3}/m_{1},\ldots,m_{n}/m_{1}\right), & (B1)\\
d=m_{2}^{2}\left(-E\right)^{-1}G\,\psi_{2}\left(m_{1}/m_{2},m_{3}/m_{2},\ldots,m_{n}/m_{2}\right), & (B2)\\
\cdots\\
d=m_{n}^{2}\left(-E\right)^{-1}G\,\psi_{n}\left(m_{1}/m_{n},m_{2}/m_{n},\ldots,m_{n-1}/m_{n}\right). & (Bn)
\end{cases}\label{eq:32-1}
\end{equation}
By (\ref{eq:sym2}), exchanging $m_{1}$ and $m_{2}$ in ($B1$) gives
\[
d=m_{2}^{2}\left(-E\right)^{-1}G\,\psi_{1}\left(m_{1}/m_{2},m_{3}/m_{2},\ldots,m_{n}/m_{2}\right)\quad(B1').
\]
and it follows from $(B1')$ and (B2) that $\psi_{1}=\psi_{2}=\psi$. 

Combining (B1) and (B2) we have
\begin{equation}
m_{1}^{2}\psi\left(m_{2}/m_{1},m_{3}/m_{1},\ldots,m_{n}/m_{1}\right)=m_{2}^{2}\psi\left(m_{1}/m_{2},m_{3}/m_{2},\ldots,m_{n}/m_{2}\right),\label{eq:321}
\end{equation}
Again setting $x_{i}=m_{i+1}/m_{1}$ for $i=1,\ldots,n-1$, (\ref{eq:321})
translates to the functional equation
\begin{equation}
\psi\left(x_{1},x_{2},\ldots,x_{n-1}\right)=x_{1}^{2}\psi\left(x_{1}^{-1},x_{1}^{-1}x_{2},\ldots,x_{1}^{-1}x_{n-1}\right).\label{eq:n-sol1}
\end{equation}
\begin{comment}
This functional equation can be derived from any two equations in
(\ref{eq:32-1}).
\end{comment}

\medskip{}

\begin{casenv}
\item (\ref{eq:n-sol1}) has solutions
\begin{equation}
\psi\left(x_{1},x_{2},\ldots,x_{n-1}\right)=C\left(\sum_{j=1}^{n-1}x_{j}+\sum_{i=1}^{n-2}\sum_{j=i+1}^{n-1}x_{i}x_{j}\right)\label{sol32-1}
\end{equation}
since
\begin{gather*}
x_{1}^{2}\left(\psi\left(x_{1}^{-1},x_{1}^{-1}x_{2},\ldots,x_{1}^{-1}x_{n-1}\right)\right)\\
=x_{1}^{2}C\left(x_{1}^{-1}+\sum_{j=2}^{n-1}x_{1}^{-1}x_{j}+\sum_{j=2}^{n-1}x_{1}^{-1}\left(x_{1}^{-1}x_{j}\right)+\sum_{i=2}^{n-2}\sum_{j=i+1}^{n-\text{1}}\left(x_{1}^{-1}x_{i}\right)\left(x_{1}^{-1}x_{j}\right)\right)\\
=C\left(x_{1}+\sum_{j=2}^{n-1}x_{1}x_{j}+\sum_{j=2}^{n-1}x_{j}+\sum_{i=2}^{n-2}\sum_{j=i+1}^{n-1}x_{i}x_{j}\right)\\
=C\left(\sum_{j=1}^{n-1}x_{j}+\sum_{i=1}^{n-2}\sum_{j=i+1}^{n-1}x_{i}x_{j}\right).
\end{gather*}
 Combining ($B1$) with (\ref{sol32-1}), we get
\begin{equation}
d=C\left(-E\right)^{-1}G\sum_{i=1}^{n-1}\sum_{j=i+1}^{n}m_{i}m_{j},\label{eq:d3-1}
\end{equation}
since
\begin{gather*}
m_{1}^{2}\left(\sum_{j=1}^{n-1}m_{j+1}/m_{1}+\sum_{i=1}^{n-2}\sum_{j=i+1}^{n-1}\left(m_{i+1}/m_{1}\right)\left(m_{j+1}/m_{1}\right)\right)\\
=\sum_{j=1}^{n-1}m_{1}m_{j+1}+\sum_{i=1}^{n-2}\sum_{j=i+1}^{n-1}m_{i+1}m_{j+1}=\sum_{i=0}^{n-2}\sum_{j=i+1}^{n-1}m_{i+1}m_{j+1}=\sum_{i=1}^{n-1}\sum_{j=i+1}^{n}m_{i}m_{j}.
\end{gather*}
Substituting (\ref{eq:d3-1}) in (\ref{eq:T3}) and setting $k_{T}^{\left(1\right)}=c_{T}C^{3}$,
we finally have
\begin{equation}
T^{2}=k_{T}^{\left(1\right)}\left(-E\right)^{-3}G^{2}\left(\sum_{i=1}^{n-1}\sum_{j=i+1}^{n}m_{i}m_{j}\right)^{3}\left(\sum_{i=1}^{n}m_{i}\right)^{-1}.\label{eq:ksun-1}
\end{equation}
 
\item (\ref{eq:n-sol1}) also has solutions
\begin{equation}
\psi\left(x_{1},x_{2},\ldots,x_{n-1}\right)=\tilde{C}\sqrt[3]{\left(\sum_{j=1}^{n-1}x_{j}^{3}+\sum_{i=1}^{n-2}\sum_{j=i+1}^{n-1}\left(x_{i}x_{j}\right)^{3}\right)}\label{sol32-2}
\end{equation}
since
\begin{gather*}
\left(x_{1}^{2}\psi\left(x_{1}^{-1},x_{1}^{-1}x_{2},\ldots,x_{1}^{-1}x_{n-1}\right)\right)^{3}\\
=x_{1}^{6}\tilde{C}^{3}\left(x_{1}^{-3}+\sum_{j=2}^{n-1}\left(x_{1}^{-1}x_{j}\right)^{3}+\sum_{j=2}^{n-1}x_{1}^{-3}\left(x_{1}^{-1}x_{j}\right)^{3}+\sum_{i=2}^{n-2}\sum_{j=i+1}^{n-\text{1}}\left(x_{1}^{-1}x_{i}\right)^{3}\left(x_{1}^{-1}x_{j}\right)^{3}\right)\\
=\tilde{C}^{3}\left(x_{1}^{3}+\sum_{j=2}^{n-1}\left(x_{1}x_{j}\right)^{3}+\sum_{j=2}^{n-1}x_{j}^{3}+\sum_{i=2}^{n-2}\sum_{j=i+1}^{n-1}\left(x_{i}x_{j}\right)^{3}\right)\\
=\tilde{C}^{3}\left(\sum_{j=1}^{n-1}x_{j}^{3}+\sum_{i=1}^{n-2}\sum_{j=i+1}^{n-1}\left(x_{i}x_{j}\right)^{3}\right).
\end{gather*}
Combining ($B1$) with (\ref{sol32-2}), we get
\begin{equation}
d^{3}=\tilde{C}^{3}\left(-E\right)^{-3}G^{3}\sum_{i=1}^{n-1}\sum_{j=i+1}^{n}\left(m_{i}m_{j}\right)^{3}\label{eq:d3}
\end{equation}
 since
\begin{gather*}
m_{1}^{6}\left(\sum_{j=1}^{n-1}\left(m_{j+1}/m_{1}\right)^{3}+\sum_{i=1}^{n-2}\sum_{j=i+1}^{n-1}\left(m_{i+1}/m_{1}\right)^{3}\left(m_{j+1}/m_{1}\right)^{3}\right)\\
=\sum_{j=1}^{n-1}\left(m_{1}m_{j+1}\right)^{3}+\sum_{i=1}^{n-2}\sum_{j=i+1}^{n-1}\left(m_{i+1}m_{j+1}\right)^{3}=\sum_{i=1}^{n-1}\sum_{j=i+1}^{n}\left(m_{i}m_{j}\right)^{3}.
\end{gather*}
Substituting (\ref{eq:d3}) in (\ref{eq:T3}) and setting $k_{T}^{\left(3\right)}=c_{T}\tilde{C}^{3}$,
we finally have
\begin{equation}
T^{2}=k_{T}^{\left(3\right)}\left(-E\right)^{-3}G^{2}\sum_{i=1}^{n-1}\sum_{j=i+1}^{n}\left(m_{i}m_{j}\right)^{3}\left(\sum_{i=1}^{n}m_{i}\right)^{-1},\label{eq:ksun}
\end{equation}
consistent with (\ref{eq:period-4-1}).
\end{casenv}

\section{Characteristic distances}

Recall that in Section \ref{sec:5} we generalized $T^{2}=4\pi^{2}d^{3}G^{-1}\left(m_{1}+m_{2}\right)^{-1}$
to
\[
T^{2}=c_{T}d{}^{3}G^{-1}\left(\sum_{i=1}^{n}m_{i}\right)^{-1}\qquad(5.6),
\]
and we generalized $T^{2}=\frac{\pi^{2}}{2}\left(-E\right)^{-3}G^{2}\left(m_{1}m_{2}\right)^{3}\left(m_{1}+m_{2}\right)^{-1}$
to 
\begin{align*}
T^{2} & =k_{T}^{\left(1\right)}\left(-E\right)^{-3}G^{2}\left(\sum_{i=1}^{n-1}\sum_{j=i+1}^{n}m_{i}m_{j}\right)^{3}\left(\sum_{i=1}^{n}m_{i}\right)^{-1}\qquad(5.12)\\
\mathrm{\mathrm{or}}\quad T^{2} & =k_{T}^{\left(3\right)}\left(-E\right)^{-3}G^{2}\sum_{i=1}^{n-1}\sum_{j=i+1}^{n}\left(m_{i}m_{j}\right)^{3}\left(\sum_{i=1}^{n}m_{i}\right)^{-1}\qquad(5.15),
\end{align*}
using (\ref{eq:T3}). In all three cases, we needed to assume that
there exists a symmetric characteristic distance $d$ for the $n$-body
system considered; we call this assumption the\emph{ characteristic
distance hypothesis}. 

This hypothesis is admittedly not compelling. Considering that already
in the three-body case the orbits can have many quite different shapes,
not only the elliptical shape found in the two-body case, one would
rather expect that $T$ would depend on one or more dimensionless
shape parameters in addition to $m_{1},\ldots,m_{n}$, $d$ and $G$.
On the other hand, the assumption that $T^{2}=\varphi{}_{T}\left(m_{1},\ldots,m_{n},d,G\right)$,
where $d$ is symmetric, makes it possible to derive predictions that
agree with data in some cases, providing indirect evidence for the
characteristic distance hypothesis.

One of the simplest generalizations of $a=\mu_{12}$ to any $n$ is
\begin{equation}
a=\frac{2}{n\left(n-1\right)}\sum_{i=1}^{n-1}\sum_{j=i+1}^{n}\mu_{ij}.\label{eq:a-new}
\end{equation}
Here, $\mu_{ij}=\frac{\left\Vert \mathbf{r}_{a}-\mathbf{0}\right\Vert +\left\Vert \mathbf{r}_{p}-\mathbf{0}\right\Vert }{2}$,
where $\mathbf{r}_{a}$ and $\mathbf{r}_{p}$ are the radius vectors
of the apoapsis and periapsis, respectively, of $p_{i}$ relative
to $p_{j},$ the radius vector of which is set to $\mathbf{0}$. As
defined by (\ref{eq:a-new}), $a$ is clearly symmetric, and for certain
simple $n$-body systems (\ref{eq:a-new}) gives the correct value
of $a$. For example, it can be shown directly from the equations
of motion that the orbital period $T$ of an equilateral triangle
three-body system is given by 
\begin{align}
T^{2} & =4\pi^{2}a^{3}G^{-1}\left(m_{1}+m_{2}+m_{3}\right)^{-1}\label{eq:eqlat}\\
 & =\frac{\pi^{2}}{2}\left(-E\right)^{-3}G^{2}\left(\left(m_{1}m_{2}\right)^{3}+\left(m_{1}m_{2}\right)^{3}+\left(m_{1}m_{2}\right)^{3}\right)\left(m_{1}+m_{2}+m_{3}\right)^{-1},\nonumber 
\end{align}
where $a$ is the length $\ell$ of the sides of the triangle. Since
$\mu_{12}=\mu_{13}=\mu_{23}=\ell$, $a$ defined by (\ref{eq:a-new})
is a characteristic distance in this case. However, numerical solutions
of equations of motion suggest that this is not true in general, although
a function $\phi_{T}'$ which also depends on a dimensionless parameter
may be based on a function $\varphi_{T}$ such that $a$ is a characteristic
distance (see Sections \ref{sec:8} and \ref{sec:9}), so the characteristic
distance hypothesis may be true in ``sufficiently many'' cases.%

\section{\label{sec:6}The constants of proportionality}

Any function obtained by dimensional analysis is unique only up to
a multiplicative constant of proportionality. Does this constant depend
on $n?$ Note that
\begin{align}
\lim_{m_{n+1}\rightarrow0} & \left(k_{T\left(n+1\right)}^{\left(1\right)}\left(-E\right)^{-3}G^{2}\left(\sum_{i=1}^{n}\sum_{j=i+1}^{n+1}m_{i}m_{j}\right)^{3}\left(\sum_{i=1}^{n+1}m_{i}\right)^{-1}\right)\label{eq:prop2}\\
 & =k_{T\left(n+1\right)}^{\left(1\right)}\left(-E\right)^{-3}G^{2}\left(\sum_{i=1}^{n-1}\sum_{j=i+1}^{n}m_{i}m_{j}\right)^{3}\left(\sum_{i=1}^{n}m_{i}\right)^{-1}.\nonumber 
\end{align}
Thus, $k_{T\left(n\right)}^{\left(1\right)}=k_{T\left(n+1\right)}^{\left(1\right)}$
and in view of (\ref{eq:period-2-2}), derived from the equations
of motion, we have $k_{T\left(2\right)}^{\left(1\right)}=\pi^{2}/2$.
Hence, $k_{T\left(n\right)}^{\left(1\right)}=\pi^{2}/2$ for all $n\geq2$
by induction.

\begin{comment}
Similarly, 
\begin{align}
\lim_{m_{n+1}\rightarrow0} & \left(k_{T\left(n+1\right)}^{\left(3\right)}\left(-E\right)^{-3}G^{2}\sum_{i=1}^{n}\sum_{j=i+1}^{n+1}\left(m_{i}m_{j}\right)^{3}\left(\sum_{i=1}^{n+1}m_{i}\right)^{-1}\right)\label{eq:kt3}\\
 & =k_{T\left(n+1\right)}^{\left(3\right)}\left(-E\right)^{-3}G^{2}\sum_{i=1}^{n-1}\sum_{j=i+1}^{n}\left(m_{i}m_{j}\right)^{3}\left(\sum_{i=1}^{n}m_{i}\right)^{-1},\nonumber 
\end{align}
so $k_{T\left(n\right)}^{\left(3\right)}=k_{T\left(n+1\right)}^{\left(3\right)}$. 
\end{comment}
Also, $k_{T\left(2\right)}^{\left(3\right)}=k_{T\left(2\right)}^{\left(1\right)}=\pi{}^{2}/2$
since
\[
\sum_{i=1}^{1}\sum_{j=i+1}^{2}\left(m_{i}m_{j}\right)^{3}=\left(\sum_{i=1}^{1}\sum_{j=i+1}^{2}m_{i}m_{j}\right)^{3}=\left(m_{1}m_{2}\right)^{3},
\]
so similarly $k_{T\left(n\right)}^{\left(3\right)}=\pi{}^{2}/2$ for
all $n\geq2$.

We have here used the fact that an $n$-body system $S$ has the same
total energy $E$ as an $n+1$-body system $S'$ which is identical
to $S$ except that it contains an additional body with mass tending
to $0$, since
\begin{align*}
\lim_{m_{n+1}\rightarrow0}\left(\sum_{i=1}^{n+1}\frac{m_{i}^{2}}{2}\left(\mathrm{\dot{q}}_{i}\right)^{2}-G\sum_{i=1}^{n}\sum_{j=i+1}^{n+1}\frac{m_{i}m_{j}}{\left\Vert \mathrm{q}_{j}-\mathrm{q}_{i}\right\Vert }\right)\\
=\sum_{i=1}^{n}\frac{m_{i}^{2}}{2}\left(\mathrm{\dot{q}}_{i}\right)^{2}-G\sum_{i=1}^{n-1}\sum_{j=i+1}^{n}\frac{m_{i}m_{j}}{\left\Vert \mathrm{q}_{j}-\mathrm{q}_{i}\right\Vert }.
\end{align*}

We cannot use the same induction approach to find the constants of
proportionality for (\ref{eq:T3}) even if $a$ given by (\ref{eq:a-new})
would be a characteristic distance for any $n$-body system, because
$a_{\left(n+1\right)}=2/\left(\left(n+1\right)n\right)\sum_{i=1}^{n}\sum_{j=i+1}^{n+1}\mu_{ij}$
does not reduce to $a_{\left(n\right)}=2/\left(n\left(n-1\right)\right)\sum_{i=1}^{n-1}\sum_{j=i+1}^{n}\mu_{ij}$
if we eliminate $p_{n+1}$. 

\section{\label{sec:8}Main results}

Recalling that $k_{T\left(n\right)}^{\left(3\right)}=k_{T\left(n\right)}^{\left(1\right)}=\pi^{2}/2$
for all $n$, we obtain
\begin{equation}
T^{2}=\frac{\pi^{2}}{2}\left(-E\right)^{-3}G^{2}\sum_{i=1}^{n-1}\sum_{j=i+1}^{n}\left(m_{i}m_{j}\right)^{3}\left(\sum_{i=1}^{n}m_{i}\right)^{-1},\label{eq:sunconj0}
\end{equation}
\begin{equation}
T^{2}=\frac{\pi^{2}}{2}\left(-E\right)^{-3}G^{2}\left(\sum_{i=1}^{n-1}\sum_{j=i+1}^{n}m_{i}m_{j}\right)^{\!3}\left(\sum_{i=1}^{n}m_{i}\right)^{\!-1}\label{eq:suncoj}
\end{equation}
from (\ref{eq:ksun}) and (\ref{eq:ksun-1}), respectively. Note that
(\ref{eq:sunconj0}) is in fact Sun's conjecture \cite{SUN1}.

If $m_{i}=m$ for $i=1,\ldots,n$ then (\ref{eq:sunconj0}) and (\ref{eq:suncoj})
simplify to
\begin{equation}
T^{2}=\frac{\pi^{2}}{4}\left(-E\right)^{-3}G^{2}m^{5}\left(n-1\right),\label{eq:equalm1}
\end{equation}
\begin{equation}
T^{2}=\frac{\pi^{2}}{16}\left(-E\right)^{-3}G^{2}m^{5}n^{2}\left(n-1\right)^{\!3},\label{eq:equalm2}
\end{equation}
respectively, since the number of terms in each of the sums $\sum_{i=1}^{n-1}\sum_{j=i+1}^{n}m_{i}m_{j}$
and $\sum_{i=1}^{n-1}\sum_{j=i+1}^{n}\left(m_{i}m_{j}\right)^{3}$
is $n\left(n-1\right)/2$, so that
\begin{gather*}
\sum_{i=1}^{n-1}\sum_{j=i+1}^{n}\left(m_{i}m_{j}\right)^{3}\left(\sum_{i=1}^{n}m_{i}\right)^{-1}=\frac{n\left(n-1\right)}{2}m^{6}\left(nm\right)^{-1}=m^{5}\left(\frac{n-1}{2}\right),\\
\left(\sum_{i=1}^{n-1}\sum_{j=i+1}^{n}m_{i}m_{j}\right)^{\!3}\left(\sum_{i=1}^{n}m_{i}\right)^{\!-1}=\left(\frac{n\left(n-1\right)}{2}m^{2}\right)^{3}\left(nm\right)^{-1}=m^{5}n^{2}\left(\frac{n-1}{2}\right)^{\!3}.
\end{gather*}

The formulas above can be compared with semi-empirical data from numerical
solutions of equations of motion in \cite{LI1,LI2}, but the comparison
is not straightforward. It turns out that the data needs to be ``normalized''
to agree with predictions from the formulas conjectured by Sun and
derived theoretically in this article.

On the one hand, (\ref{eq:sunconj0}) gives the actual period for
many $n$-body systems with a central configuration. For example,
recall that for an equilateral-triangle three-body system $T$ is
given by (\ref{eq:eqlat}).

On the other hand, however, orbital $n$-body systems typically have
more complicated orbits that take longer time to complete than that
given by (\ref{eq:sunconj0}). There are indications that orbital
periods are, at least approximately, integer multiples of the periods
given by (\ref{eq:sunconj0}), as if an orbit could be broken down
into ``loops'' with equal periods, each given by (\ref{eq:sunconj0}).
Remarkably, Montgomery \cite{MONT} found that the complexity of an
orbit in an orbital $n$-body system increased with the length $L_{\mathrm{f}}$
of a corresponding word in a certain free group, and Li and Liao \cite{LI1}
then showed that in a three-body system where $m_{1}=m_{2}=m_{3}=G=1$
we have, approximately, $\bar{T}\propto\left(-E\right)^{-3/2}$, where
$\bar{T}$ is the ``loop period'' $T/L_{\mathrm{f}}$. Sun's conjecture
is supported by data from \cite{LI1,LI2} only if $T$ is identified
with $\bar{T}$. When comparing data with predictions below, $T$
will be replaced by $\bar{T}=T/L_{\mathrm{f}}$, where $L_{\mathrm{f}}$
may also equal $1$. 

If $n=3$ and we set $m_{1}=m_{2}=m_{3}=G=1$ then (\ref{eq:equalm1})
and (\ref{eq:equalm2}) simplify further to $\bar{T}^{2}=\frac{\pi^{2}}{2}\left(-E\right)^{-3}$
and $\bar{T}^{2}=\frac{9\pi^{2}}{2}\left(-E\right)^{-3}$, respectively,
so that $\bar{T}\left(-E\right)^{3/2}=\pi/\sqrt{2}\approx2.221$ and
$\bar{T}\left(-E\right)^{3/2}=3\pi/\sqrt{2}\approx6.664$, respectively.
Li and Liao \cite{LI1} obtain $\bar{T}\left(-E\right)^{3/2}=2.433\pm0.075$
by solving equations of motion numerically.\footnote{In \cite{LI1}, $\bar{T}$ is the mean of all $\widehat{T}/L_{\mathrm{f}}$,
where $\widehat{T}$ is the period and $L_{\mathrm{f}}$ the word
length of an individual computer-calculated orbit. Similarly, the
values of $\bar{T}$ from \cite{LI2} are the means of $\widehat{T}/L_{\mathrm{f}}$
over all orbits corresponding to $m_{3}=0.5,0.75,2,4,5,8,10$, respectively.} 

If we instead set $m_{1}=m_{2}=G=1$ and let $m_{3}$ vary then (\ref{eq:sunconj0})
gives 
\[
\bar{T}\left(-E\right)^{3/2}=\sqrt{\frac{\pi^{2}}{2}\sum\nolimits_{i=1}^{2}\sum\nolimits_{j=i+1}^{3}\left(m_{i}m_{j}\right)^{3}\left(\sum\nolimits_{i=1}^{3}m_{i}\right)^{-1}}\simeq\pi m_{3}\approx3.142m_{3}
\]
as $m_{3}\rightarrow\infty$, while Li et al. \cite{LI2} obtain $\bar{T}\left(-E\right)^{3/2}\approx3.074m_{3}-0.617$.
Sun shows \cite[Figure 4]{SUN1} that the values obtained from (\ref{eq:sunconj0})
in general agree well with the data in \cite{LI2} while avoiding
the anomaly that for the two-body system obtained by setting $m_{3}=0$
the formula in \cite{LI2} yields $\bar{T}<0$.

Note, however, that the quantity $\bar{T}^{2}\left(-E\right)^{3}$
is not dimensionless, so its measure depends on a choice of units
of measurement \cite{JON3}. Setting $m_{i}=G=1$ amounts to choosing
a unit $u_{\mathsf{M}}$ for $\mathsf{M}$ and a unit $u_{\left[G\right]}$
for $\left[G\right]$ such that the measures of $m_{i}$ and $G$
both equal 1, and as $\mathsf{T}^{2}\mathsf{E}^{3}=\mathsf{M}{}^{5}\left[G\right]^{2}$
the quantity $u_{\mathsf{M}}^{5}u_{\left[G\right]}^{2}$ is the unit
for $\mathsf{T}^{2}\mathsf{E}^{3}$ such that numerically $\bar{T}^{2}\left(-E\right)^{3}\approx2.433^{2}$
in \cite{LI1}.%
\begin{comment}
choice of units for the dimensions $\mathsf{M}$ and $\left[G\right]$
(cf. \cite{JON3,RAP}),

Note, however, that the quantity $\bar{T}^{2}\left(-E\right)^{3}$
is not dimensionless, so as a scalar it depends on a choice of units
of measurement \cite{JON3}. Setting $m_{i}=G=1$ amounts to a choice
of units for the dimensions $\mathsf{M}$ and $\left[G\right]$ (cf.
\cite{JON3,RAP}), and as $\mathsf{T}^{2}\mathsf{E}^{3}=\mathsf{M}{}^{5}\left[G\right]^{2}$
the choice of units for $\mathsf{M}$ and $\left[G\right]$ determines
a consistent choice of unit for $\mathsf{T}^{2}\mathsf{E}^{3}$. 
\end{comment}
{} Thus, the value of $\bar{T}\left(-E\right)^{3/2}$ depends on a choice
of unit implied by the assumption that $m_{1}=m_{2}=m_{3}=G=1$.

In contrast to (\ref{eq:sunconj0}), the analytic solution (\ref{eq:suncoj}),
although also dimensionally homogeneous, did not agree with the numerical
solutions in \cite{LI1,LI2}. However, Semay \cite[eq. 9]{SEM1} suggested
that a certain quantum-theoretical analogue $T_{q}$ of $T$ for a
system of $n$ identical quantum particles is given by (\ref{eq:equalm2})
with $T$ replaced by $T_{q}$ (see also \cite{SEM2}). Sun \cite{SUN2}
then suggested that in a more general system of $n$ quantum particles
$T_{q}$ is given by (\ref{eq:suncoj}) with $T$ replaced by $T_{q}$,
noting that (\ref{eq:equalm2}) follows from (\ref{eq:suncoj}) under
the assumption that all particle masses are equal. 

We conclude that the agreement between Sun's formula and the data
obtained by Li et al. as well as that between Sun's and Semay's formulas
are not coincidental. In particular, the ultimate reason why the expression
(\ref{eq:equalm1}) for $T$ in the classical case and (\ref{eq:equalm2})
for $T_{q}$ in the quantum-theoretical case differ only by a factor
of 
\[
\sqrt{n^{2}\left(n-1\right)^{3}/16}\:\big/\sqrt{\left(n-1\right)/4}=n\left(n-1\right)/2
\]
is the fact that both (\ref{sol32-1}) and (\ref{sol32-2}) are solutions
of the functional equation (\ref{eq:n-sol1}).

\section{\label{sec:9}Conclusions about uniqueness}

Let us return to the question of the uniqueness of a generalization
of 
\[
T^{2}=\frac{\pi^{2}}{2}\left(-E\right)^{-3}G^{2}\left(m_{1}m_{2}\right)^{3}\left(m_{1}+m_{2}\right)^{-1}\qquad(1.5),
\]
derived from the equations of motion, to an equation of the form
\begin{equation}
T^{\mathfrak{J}}=\phi_{T}\left(m_{1},\ldots,m_{n},E,G\right).\label{eq:gen}
\end{equation}
There are two basic requirements on such generalizations.

First, the equation (\ref{eq:gen}) must be dimensionally homogeneous,
since otherwise it would not be physically meaningful. Second, $\phi_{T}$
must be symmetric in $m_{1},\ldots,m_{n}$. In Section \ref{subsec:21},
we proved this formally, using the fact that the permutations 
\[
\begin{cases}
m_{i}\mapsto m_{\sigma\left(i\right)},\quad\mathbf{q}_{i}\mapsto\mathbf{q}_{\sigma\left(i\right)} & \left(i=1,\ldots,n\right)\end{cases}
\]
that correspond to a relabeling of $p_{1},\ldots,p_{n}$ maps the
related system of equations of motion and initial conditions to itself.
Informally, the labeling of the bodies in an $n$-body system is just
a reference frame, and as always the physical reality remains unchanged
when the reference frame is changed, implying the required symmetries.

Unfortunately, these two criteria do not sufficiently restrict the
range of possible equations. For example, 
\begin{gather*}
T^{2}=k_{1}\left(-E\right)^{-3}G^{2}\frac{\left(m_{1}m_{2}\right)^{3}}{\sqrt{m_{1}m_{2}}},\qquad T^{2}=k_{2}\left(-E\right)^{-3}G^{2}\left(m_{1}^{5}+m_{2}^{5}\right),
\end{gather*}
and corresponding equations for $n>2$, satisfy both conditions, but
do not have the same form as (\ref{eq:period-2-2}), and do not fit
the data.

However, we have assumed above that there exists a symmetric distance
$d$ and a function $\varphi_{T}$ such that $T^{\mathfrak{K}}=\varphi_{T}\left(m_{1},\ldots,m_{n},d,G\right)$,
as well as a function $\phi_{d}$ such that $d^{\mathfrak{L}}=\phi_{d}\left(m_{1},\ldots,m_{n},E,G\right)$.
Using these assumptions, as we did in Sections \ref{sec:3}\textendash \ref{sec:5},
we restrict the range of candidate equations further, in particular
not encountering the two displayed equations in the preceding paragraph. 

Specifically, there is one candidate equation of the form (\ref{eq:gen})
for every solution of the functional equation
\[
\psi\left(x_{1},x_{2},\ldots,x_{n-1}\right)=x_{1}^{2}\psi\left(x_{1}^{-1},x_{1}^{-1}x_{2},\ldots,x_{1}^{-1}x_{n-1}\right)\qquad(5.9)
\]
in Section \ref{sec:5}, and it turns out that 
\begin{equation}
\psi_{\left(q\right)}\left(x_{1},x_{2},\ldots,x_{n-1}\right)=k_{d}^{\left(q\right)}\left(\sum_{i=1}^{n-1}x_{i}^{q}+\sum_{i=1}^{n-2}\sum_{j=i+1}^{n-1}\left(x_{i}x_{j}\right)^{q}\right)^{1/q}\label{eq:pe}
\end{equation}
is a solution of (5.9) for every positive integer $q$. Equations
(\ref{eq:sunconj0}) and (\ref{eq:suncoj}) above correspond to $q=3$
and $q=1$, respectively.

This information can be used in more than one way. On the one hand,
we may ask if there is something special with the numbers $1$ and
$3$ here, so that only (\ref{eq:sunconj0}) and (\ref{eq:suncoj})
are non-spurious solutions. First, it is shown in \cite[Appendix A]{JON1})
that a functional equation of the form $\psi\left(x\right)=x\,\psi\left(x^{-1}\right)$
has a unique \emph{analytical} solution, namely $\psi\left(x\right)=k\left(1+x\right)$.
Similarly, setting $q=1$ we obtain the unique analytical solution
$\psi_{\left(1\right)}$ of (5.9). Second, the function $\psi_{\left(3\right)}^{3}$
given by
\[
\psi_{\left(3\right)}^{3}\left(x_{1},x_{2},\ldots,x_{n-1}\right)=k_{d}^{\left(3\right)}\left(\sum_{i=1}^{n-1}x_{i}^{3}+\sum_{i=1}^{n-2}\sum_{j=i+1}^{n-1}\left(x_{i}x_{j}\right)^{3}\right)
\]
is analytic, $d$ appears with the exponent 3 in (\ref{eq:T3}) and
we obtain the same result if we substitute $\psi_{\left(3\right)}^{3}\left(x_{1},x_{2},\ldots,x_{n-1}\right)$
for $d^{3}$ as if we substitute $\psi_{\left(3\right)}\left(x_{1},x_{2},\ldots,x_{n-1}\right)$
for $d$. Thus, it may be argued that $q=1$ and $q=3$ are special,
privileged cases.

On the other hand, there may nevertheless be additional non-spurious
solutions to (5.9), corresponding to different generalizations of
(\ref{eq:period-2-2}) to the case $n>2$. Many topologically distinguishable
types of orbital trajectories exist \cite{MONT,DIM}, and it is possible
that solutions of (5.9) for different values of $q$ correspond to
different families of $n$-body systems.

Note furthermore that the functional equation
\[
\psi\left(x_{1},x_{2},\ldots,x_{n-1}\right)=x_{1}^{-1}\psi\left(x_{1}^{-1},x_{1}^{-1}x_{2},\ldots,x_{1}^{-1}x_{n-1}\right)\qquad(5.4),
\]
also used in the derivation of generalizations of (\ref{eq:period-2-2}),
similarly has a solution 
\begin{equation}
\psi_{\left(q\right)}\left(x_{1},x_{2},\ldots,x_{n-1}\right)=c_{T}^{\left(q\right)}\left(1+\sum_{i=1}^{n-1}x_{i}^{q}\right)^{-1/q}\label{eq:pt}
\end{equation}
for every positive integer $q$. The solution used in Sections \ref{sec:3}\textendash \ref{sec:5}
is the one for $q=1$. 

Any particular solution of (5.4) can be combined with any particular
solution of (5.9) into a generalization of (\ref{eq:period-2-2}).
Thus, the possible solutions may form a kind of two-dimensional space,
suggesting an alternative approach to the great variability of orbital
periods for $n>2$ touched on in Section 8. %
\begin{comment}
There is of course another, more straight-forward way to eliminate
spurious equations, namely to compare their predictions to numerical
solutions of equations of motion. Such a comparison was made by Sun
\cite{SUN1} for $n=3$, but in practice as well as in principle this
approach cannot be extended to all $n$. Thus, theoretical considerations
as formulated in this section are also important.
\end{comment}

For completeness, it should also be mentioned that as suggested by
the examples in \cite[pp. 16--18]{JON1} there are more solutions
to (5.4) and (5.9) than those considered above. Whether or not these
more complicated solutions are relevant in the context of $n$-body
systems remains to be researched.

We conclude that the variation in $T$ that is not attributable to
the parameters $m_{1},\ldots,m_{n}$ and $a$ or $E$ can be attributed,
at least partly, to the invocation of different solutions to the corresponding
functional equation. For example, the difference between (\ref{eq:sunconj0})
and (\ref{eq:suncoj}) can be explained in this way. Within limits,
the non-uniqueness of solutions of functional equations obtained by
augmented dimensional analysis may thus be a strength rather than
a weakness. 

But as mentioned earlier, remaining variation in $T$ can also be
accounted for by one or more additional dimensionless parameters.
The data from \cite{LI1,LI2} do indeed suggest that $T$ depends
on the integer $L_{\mathrm{f}}$, so that
\[
T^{2}=\phi_{T}'\left(m_{1},\ldots,m_{n},E,L_{\mathrm{f}},G\right).
\]
 Specifically, the data suggest that $\phi_{T}'$ admits the factorization
$\phi_{T}'=L_{\mathrm{f}}^{2}\phi_{T}$, that is, 
\[
T^{2}=L_{\mathrm{f}}^{2}\phi_{T}\left(m_{1},\ldots,m_{n},E,G\right).
\]
Hence, given $L_{\mathrm{f}}$, $\phi_{T}'$ can be derived from $\varphi{}_{T}$
and $\phi_{d}$, where 
\[
T^{2}=\varphi_{T}\left(m_{1},\ldots,m_{n},a,G\right),
\]
 so that $a$ is a characteristic distance.

\end{document}